\documentclass[aps,pre,groupedaddress,twocolumn,amsmath,showpacs]{revtex4}
    \usepackage{bm}
    \usepackage{graphicx}
\newcommand\beq{\begin{equation}}
\newcommand\eeq{\end{equation}}
\newcommand\beqa{\begin{eqnarray}}
\newcommand\eeqa{\end{eqnarray}}
\newcommand{\nn}{\nonumber\\}

\begin{document}
\title{Exact bulk correlation functions in one-dimensional nonadditive hard-core mixtures}
\author{Andr\'es Santos}
\email{andres@unex.es} \homepage{http://www.unex.es/fisteor/andres/}
\affiliation{Departamento de F\'{\i}sica, Universidad de
Extremadura, E-06071 Badajoz, Spain}
\date{\today}

\begin{abstract}
In a recent paper [Phys.\ Rev.\ E \textbf{76}, 031202 (2007)],
Schmidt has proposed a Fundamental Measure Density Functional Theory
for one-dimensional nonadditive hard-rod fluid mixtures and has
compared its predictions for the bulk structural properties with
Monte Carlo simulations. The aim of this Brief Report  is to recall
that the problem admits an exact solution in the bulk, which is
briefly summarized in a self-contained way.
\end{abstract}

\maketitle

Perhaps the most successful class of density functional theories are
based on Rosenfeld's Fundamental Measure Theory (FMT) \cite{R89}. In
a recent paper \cite{M07}, Schmidt has proposed a FMT for the excess
free energy of inhomogeneous one-dimensional nonadditive hard-rod
fluid mixtures. As a test of the theory, the FMT predictions for the
pair correlation functions in the bulk region are compared with
Monte Carlo simulations, a general good agreement being found. On
the other hand, notwithstanding the merits of the FMT constructed in
Ref.\ \cite{M07}, it presents some limitations that become more
important as the density and/or the nonadditivity increase. For
instance, it yields non-zero values of the pair correlation
functions inside the core and predicts a spurious demixing
transition.

It seems to have been overlooked in Ref.\ \cite{M07} the fact that
the one-dimensional nonadditive hard-rod problem admits an
\emph{exact} solution in the bulk. Actually, any one-dimensional
homogeneous system is exactly solvable, provided that every particle
interacts only with its nearest neighbors \cite{SZK53,LZ71,HC04}.
The aim of this Brief Report is to fill the gap in Ref.\ \cite{M07}
by presenting a brief and self-contained summary of the exact
solution, particularizing to binary nonadditive mixtures, and
comparing with the bulk FMT predictions for one of the cases
considered in Ref.\ \cite{M07}.

Let us consider an $m$-component one-dimensional fluid mixture with
constant (bulk) number densities $\{\rho_i;i=1,\ldots,m\}$ and
interaction potentials $\phi_{ij}(x)=\phi_{ij}(-x)$ acting only on
nearest neighbors.
 Given a particle of species
 $i$ at the origin, the probability that its $\ell$th neighbor
 belongs to species $j$ and is located at a point between $x$ and $x+dx$ is given by
 $p^{(\ell)}_{ij}(x)dx$,
what defines the (conditional) probability density distribution
$p^{(\ell)}_{ij}(x)$. In particular, $p^{(1)}_{ij}(x)$ is the
nearest-neighbor distribution. The distributions
$p^{(\ell)}_{ij}(x)$ verify the normalization condition
\beq
\sum_{j=1}^m\int_0^\infty dx\, p^{(\ell)}_{ij}(x)=1
\label{n5}
\eeq
and obey the recurrence relation
\beq
p^{(\ell)}_{ij}(x)=\sum_{k=1}^m\int_0^x dx' \,
p^{(\ell-1)}_{ik}(x')p^{(1)}_{kj}(x-x').
\label{n7}
\eeq
Its solution in Laplace space is
\beq
\mathsf{P}^{(\ell)}(s)=\left[\mathsf{P}^{(1)}(s)\right]^\ell,
\label{n10}
\eeq
where $\mathsf{P}^{(\ell)}(s)$ is the $m\times m$ matrix whose
elements $P^{(\ell)}_{ij}(s)$ are the  the Laplace transforms of
$p^{(\ell)}_{ij}(x)$.

The total  probability density of finding a particle of species $j$,
given that a particle of species $i$ is at the origin, is obtained
as
\beq
\rho_j g_{ij}(x)=p_{ij}(x)=\sum_{\ell=1}^\infty p_{ij}^{(\ell)}(x),
\label{n5.2}
\eeq
where  $g_{ij}(x)$ is the pair correlation function. In Laplace
space,
\beq
G_{ij}(s)=\frac{1}{\rho_j}P_{ij}(s),\quad
\mathsf{P}(s)=\mathsf{P}^{(1)}(s)\cdot
\left[\mathsf{I}-\mathsf{P}^{(1)}(s)\right]^{-1},
\label{n12}
\eeq
where use has been made of Eq.\ \eqref{n10}. Therefore, the
knowledge of the nearest-neighbor distributions
$\{p_{ij}^{(1)}(x)\}$ suffices to obtain the pair correlation
functions $\{g_{ij}(x)\}$. Note that the Fourier transform
$\widetilde{h}_{ij}(k)$ of the total correlation function
$h_{ij}(x)\equiv g_{ij}(x)-1$ is simply related to the Laplace
transform $G_{ij}(s)$ of $g_{ij}(x)$ by
$\widetilde{h}_{ij}(k)=G_{ij}(\imath k)+G_{ij}(-\imath k)$, where
$\imath$ is the imaginary unit.

It can be proven that  the nearest-neighbor distribution possesses
the following explicit form \cite{LZ71,HC04}:
\beq
p_{ij}^{(1)}(x)=\rho_j K_{ij}e^{-\beta\phi_{ij}(x)}e^{-\xi x},
\label{n32b}
\eeq
where $\beta=1/k_BT$ and $\xi=\beta p$, $k_B$,  $T$, and $p$ being
the Boltzmann constant, the temperature, and the pressure,
respectively.  The Laplace transform of Eq.\ \eqref{n32b} is
\beq
P_{ij}^{(1)}(s)=\rho_j K_{ij}\Omega_{ij}(s+\xi),
\label{n35}
\eeq
where $\Omega_{ij}(s)$ denotes the Laplace transform of
$e^{-\beta\phi_{ij}(x)}$.

To close the problem, one needs to determine the amplitudes
$K_{ij}=K_{ji}$ and the damping coefficient $\xi$. A convenient way
of doing so is  by enforcing basic consistency conditions. Note
first that the normalization condition \eqref{n5} for $\ell=1$ is
equivalent to
\beq
\sum_{j=1}^m{P}_{ij}^{(1)}(0)=1.
\label{n18}
\eeq
 Next,
since $\lim_{x\to\infty}g_{ij}(x)= 1$,  one must have
\beq
\lim_{s\to 0}sG_{ij}(s)=1.
\label{n19}
\eeq
A subtler consistency condition   \cite{LZ71} dictates that
$\lim_{x\to\infty}{p_{ij}^{(1)}(x)}/{p_{ik}^{(1)}(x)}$  must be
independent of the choice of species $i$. {}From Eq.\ \eqref{n32b}
this implies that
\beq
\frac{K_{ij}}{K_{ik}}=\text{independent of $i$}.
\label{nsymm_bis}
\eeq

Equations \eqref{n18}--\eqref{nsymm_bis} are sufficient to obtain
$K_{ij}$ and $\xi$. To be more specific, let us consider the case of
a binary mixture ($m=2$). Thus,  Eq.\ \eqref{n12} yields
\beq
G_{11}(s)=\frac{Q_{11}(s)\left[1-Q_{22}(s)\right]+Q_{12}^2(s)}{\rho_1
D(s)},
\label{22}
\eeq
\beq
G_{22}(s)=\frac{Q_{22}(s)\left[1-Q_{11}(s)\right]+Q_{12}^2(s)}{\rho_2
D(s)},
\label{23}
\eeq
\beq
G_{12}(s)=\frac{Q_{12}(s)}{\sqrt{\rho_1\rho_2}D(s)},
\label{24}
\eeq
where
\beq
Q_{ij}(s)\equiv \sqrt{\rho_i/\rho_j}P_{ij}^{(1)}(s)=
\sqrt{\rho_i\rho_j} K_{ij}\Omega_{ij}(s+\xi),
\label{35}
\eeq
\beq
D(s)\equiv\left[1-Q_{11}(s)\right]\left[1-Q_{22}(s)\right]-Q_{12}^2(s).
\label{26}
\eeq
The behavior of $Q_{ij}(s)$ for small $s$ is
\beq
Q_{ij}(s)=\sqrt{\rho_i\rho_j}
K_{ij}\left[\Omega_{ij}(\xi)+\Omega_{ij}'(\xi)
s+\mathcal{O}(s^2)\right],
\label{32}
\eeq
where $\Omega_{ij}'(s)$ is the first derivative of $\Omega_{ij}(s)$.
Application of Eq.\ \eqref{n18} yields
\beq
K_{11}=\frac{1-\rho_2K_{12}\Omega_{12}(\xi)}{\rho_1\Omega_{11}(\xi)},
\label{33}
\eeq
\beq
K_{22}=\frac{1-\rho_1K_{12}\Omega_{12}(\xi)}{\rho_2\Omega_{22}(\xi)}.
\label{34}
\eeq
Next, Eq.\ \eqref{n19}  implies
\beq
\rho_1^2 K_{11}\Omega_{11}'(\xi)+\rho_2^2
K_{22}\Omega_{22}'(\xi)+2\rho_1\rho_2 K_{12}\Omega_{12}'(\xi)=-1.
\label{31}
\eeq
Finally, Eq.\ \eqref{nsymm_bis} becomes
\beq
K_{11}K_{22}=K_{12}^2.
\label{30}
\eeq
Equations \eqref{33}--\eqref{30} constitute a set of four
independent equations whose solution gives $K_{11}$, $K_{12}$,
$K_{22}$, and $\xi$. Inserting Eqs.\ \eqref{33} and \eqref{34} into
Eqs.\ \eqref{31} and \eqref{30} one gets
\beq
K_{12}=\frac{1}{\rho_1\rho_2\Omega_{12}(\xi)}\frac{1+\rho_1L_{11}(\xi)+\rho_2L_{22}(\xi)}{L_{11}(\xi)+L_{22}(\xi)
-2L_{12}(\xi)},
\label{38}
\eeq
\beq
1-\rho
K_{12}\Omega_{12}(\xi)+\rho_1\rho_2\left[\Omega_{12}^2(\xi)-{\Omega_{11}(\xi)\Omega_{22}(\xi)}\right]
K_{12}^2=0,
\label{43}
\eeq
where  we have called $L_{ij}(s)\equiv
\Omega_{ij}'(s)/\Omega_{ij}(s)$ and $\rho=\rho_1+\rho_2$ is the
total density. Substitution of Eq.\ \eqref{38} into Eq.\ \eqref{43}
yields a single  equation for $\xi$, which in general is
transcendental. Once solved, the coefficients $K_{ij}$ are obtained
from Eqs.\ \eqref{33}, \eqref{34}, and \eqref{38}. The exact pair
correlation functions are then entirely determined in Laplace space
through Eqs.\  \eqref{22}--\eqref{26}.

\begin{figure}
\includegraphics[width=\columnwidth]{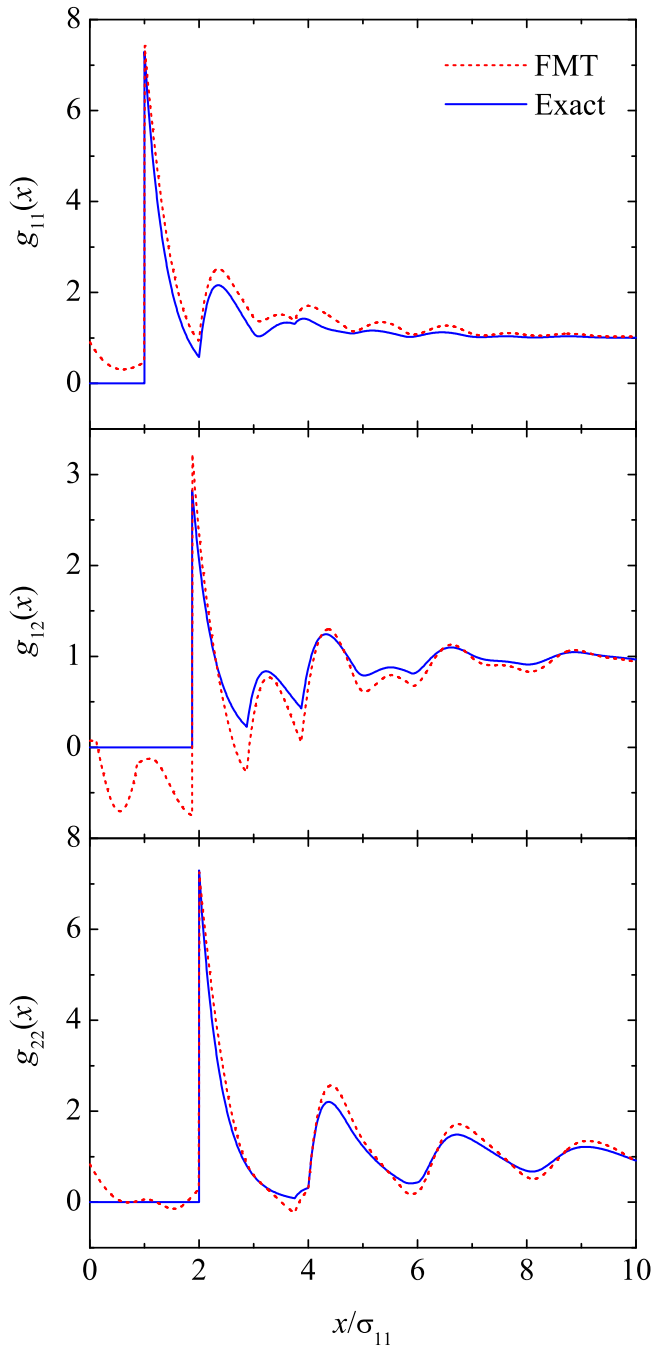}
\caption{(Color online) Bulk pair correlation functions $g_{ij}(x)$
for a one-dimensional binary hard-rod mixture with
$\sigma_{22}/\sigma_{11}=2$, $\sigma_{12}/\sigma_{11}=15/8$, and
$\rho_1=\rho_2=\sigma_{11}^{-1}/4$. The solid lines are the exact
results and the dashed lines are the FMT predictions of Ref.\
\protect\cite{M07}.}
\label{figure}
\end{figure}

In the particular case of nonadditive hard rods, one has
$e^{-\beta\phi_{ij}(x)}=\Theta(x-\sigma_{ij})$, where $\Theta(x)$ is
Heaviside's step function, so that
\beq
\Omega_{ij}(s)=\frac{e^{-\sigma_{ij}s}}{s},\quad
L_{ij}(s)=-\sigma_{ij}-s^{-1},
\label{44}
\eeq
\beq
Q_{ij}(s)=\sqrt{\rho_i\rho_j}
K_{ij}\frac{e^{-\sigma_{ij}(s+\xi)}}{s+\xi}.
\label{n35bis}
\eeq
The constraint to nearest-neighbor interactions implies that
$\sigma_{ij}\leq\sigma_{ik}+\sigma_{jk}$ for all $\{i,j,k\}$. In the
binary case this amounts to
$2\sigma_{12}>\text{max}(\sigma_{11},\sigma_{22})$. The recipe
described by Eqs.\ \eqref{33}, \eqref{34}, \eqref{38},  \eqref{43},
and \eqref{44}  for the thermodynamic quantity $\xi=\beta p$ and the
amplitudes $K_{ij}$, and by Eqs.\ \eqref{22}--\eqref{26} and
\eqref{n35bis} for the structural quantities $G_{ij}(s)$ are easy to
implement. In order to go back to real space and obtain the pair
correlation functions $g_{ij}(x)$ one can use any of the efficient
numerical schemes described in Ref.\ \cite{AW92}. On the other hand,
the simplicity of Eq.\ \eqref{n35bis} allows one to get a fully
analytical representation. Note first that
\beq
\frac{1}{D(s)}=\sum_{m=0}^\infty
\left[Q_{11}(s)+Q_{22}(s)+Q_{12}^2(s)-Q_{11}(s)Q_{22}(s)\right]^m.
\label{45}
\eeq
When Eq.\ \eqref{45} is inserted into Eqs.\ \eqref{22}--\eqref{24},
one can express $G_{ij}(s)$ as linear combinations of terms of the
form
\beqa
Q_{11}^{n_{11}}(s)Q_{22}^{n_{22}}(s)Q_{12}^{n_{12}}(s)&=&\frac{e^{-a(s+\xi)}}{(s+\xi)^n}\left(\rho_1
K_{11}\right)^{n_{11}+n_{12}/2} \nn &&\times \left(\rho_2
K_{22}\right)^{n_{22}+n_{12}/2},
\label{47}
\eeqa
where $a\equiv n_{11}
\sigma_{11}+n_{22}\sigma_{22}+n_{12}\sigma_{22}$ and $n\equiv
n_{11}+n_{22}+n_{12}$. The inverse Laplace transforms
$g_{ij}(x)=\mathcal{L}^{-1}\left[G_{ij}(s)\right]$ are readily
evaluated by using the property
\beq
\mathcal{L}^{-1}\left[\frac{e^{-a(s+\xi)}}{(s+\xi)^n}\right]=e^{-\xi
x}\frac{(x-a)^{n-1}}{(n-1)!}\Theta(x-a).
\label{46}
\eeq
 It is important to realize that if
one is interested in distances $x$ smaller than a certain value $R$,
only a \emph{finite} numbers of terms contribute to $g_{ij}(x)$,
namely those with $\{n_{11},n_{22},n_{12}\}$ such that $n_{11}
\sigma_{11}+n_{22}\sigma_{22}+n_{12}\sigma_{22}<R$. In particular,
for the most nonadditive case considered in Ref.\ \cite{M07}, i.e.,
$\sigma_{22}/\sigma_{11}=2$ and $\sigma_{12}/\sigma_{11}=15/8$, only
those terms satisfying $8n_{11}+16n_{22}+15n_{12}<80$ are needed for
$x<10\sigma_{11}$. Moreover,
$g_{ij}(x)=\rho_j^{-1}p_{ij}^{(1)}(x)=K_{ij}e^{-\xi x}$ in the first
shell, i.e., for $\sigma_{ij}<x<\sigma_{ij}+\Delta_{ij}$, where
$\Delta_{11}=\text{min}(\sigma_{11},2\sigma_{12}-\sigma_{11})$,
$\Delta_{22}=\text{min}(\sigma_{22},2\sigma_{12}-\sigma_{22})$, and
$\Delta_{12}=\text{min}(\sigma_{11},\sigma_{22})$.

Let us consider a specific  system with $\sigma_{22}/\sigma_{11}=2$,
$\sigma_{12}/\sigma_{11}=15/8$, and
$\rho_1=\rho_2=\sigma_{11}^{-1}/4$. The corresponding solution of
the transcendental equation for $\xi$ is $\xi\simeq 2.52964
\sigma_{11}^{-1}$, so that $\beta p/\rho\simeq 5.05927$. The
numerical values of the amplitudes $K_{ij}$ and the contact values
$g_{ij}(\sigma_{ij}^+)$ are $K_{11}\simeq 91.5298$, $K_{22}\simeq
1148.60$, $K_{22}\simeq 324.24$,
$g_{11}(\sigma_{11}^+)=g_{22}(\sigma_{22}^+)\simeq 7.29382$, and
$g_{12}(\sigma_{12}^+)\simeq 2.82473$. The property
$g_{11}(\sigma_{11}^+)=g_{22}(\sigma_{22}^+)$ is common to all the
equimolar cases ($\rho_1=\rho_2$), since then Eqs.\ \eqref{33} and
\eqref{34} imply that
$K_{11}\Omega_{11}(\xi)=K_{22}\Omega_{22}(\xi)$. Figure \ref{figure}
compares the three exact bulk correlation functions $g_{ij}(x)$ with
those predicted by the FMT proposed in Ref.\ \cite{M07}. The
discrepancies are similar to those found in Ref.\ \cite{M07} between
Monte Carlo simulations and FMT.

It must be emphasized that the scheme \eqref{n12}--\eqref{nsymm_bis}
provides the exact \emph{bulk} correlation functions for a
one-dimensional  mixture in the \emph{absence} of external fields.
The more general problem addressed in Ref.\ \cite{M07}, namely the
excess free energy as a functional of the inhomogeneous densities,
is much more complicated and, to the best of my knowledge, its exact
solution is not known. On the other hand,  the exact density
profiles $\rho_j(x)$ induced by external potentials $V_j(x)$ can be
obtained  under certain conditions. The trick consists of assuming
that one of the species (here labeled as $i=0$) has a vanishing
concentration ($\rho_0=0$) and interacts with the other species via
the potentials $\phi_{0j}(x)=V_j(x)$. The knowledge of the bulk
correlation functions $g_{ij}(x)$   (with
$\rho_j\to\rho_j^{\text{bulk}}$) can then be exploited to get
$\rho_j(x)=\rho_j^{\text{bulk}}g_{0j}(x)$. The important limitation,
however, is that $V_j(x)$ must represent the potential exerted by a
wall that acts only on its nearest particles.

To conclude, it is expected that the exact solutions for
one-dimensional homogeneous systems derived elsewhere
\cite{SZK53,LZ71,HC04} and summarized in this paper can be useful as
benchmarks to construct, test, and refine approximate theories like
the FMT of Ref.\ \cite{M07}. This would allow one to gain some
illuminating insight into the subtleties and difficulties of the
problem of interest, which can be helpful in its extension to the
more realistic case of three-dimensional systems.

\begin{acknowledgments}

I am grateful to M. Schmidt for kindly providing  the FMT values
represented in Fig.\ \ref{figure}. This work has been supported by
the Ministerio de Educaci\'on y Ciencia (Spain) through Grant No.\
FIS2007--60977 (partially financed by FEDER funds) and by the Junta
de Extremadura through Grant No.\ GRU07046.

\end{acknowledgments}

\end{document}